\documentclass[longauth,structabstract,letter]{aa} 
\usepackage{txfonts}
\usepackage{epsfig}
\usepackage{subfigure}

\usepackage{times}
\usepackage{natbib}
\usepackage{rotating}
\usepackage{lscape}
\bibpunct{(}{)}{;}{a}{}{,}

\begin{document}

\title{CLASH-VLT: spectroscopic confirmation of a z=6.11 quintuply lensed galaxy 
in the Frontier Fields Cluster RXC J2248.7-4431
\thanks{This work is based on data collected at ESO VLT (prog.ID 186.A-0798) 
and at NASA HST.}}

\author{I. Balestra\inst{\ref{ABi},\ref{AMe}}
\and E. Vanzella\inst{\ref{SEt}}
\and P. Rosati\inst{\ref{PRo}}
\and A. Monna\inst{\ref{SSe1}}
\and C. Grillo\inst{\ref{CGr}}
\and M. Nonino\inst{\ref{ABi}} 
\and A. Mercurio\inst{\ref{AMe}} 
\and A. Biviano\inst{\ref{ABi}}
\and L. Bradley\inst{\ref{MPo}}
\and D. Coe\inst{\ref{MPo}}
\and A. Fritz\inst{\ref{MSc}}
\and M. Postman\inst{\ref{MPo}}
\and S. Seitz\inst{\ref{SSe1},\ref{SSe2}}
\and M. Scodeggio\inst{\ref{MSc}}
\and P. Tozzi\inst{\ref{PTo}}
\and W. Zheng\inst{\ref{DLe}}
\and B. Ziegler\inst{\ref{OCz}}
\and A. Zitrin\inst{\ref{MBa}}
\and M. Annunziatella\inst{\ref{MGi},\ref{ABi}}
\and M. Bartelmann\inst{\ref{MBa}}
\and N. Benitez\inst{\ref{NBe}}
\and T. Broadhurst\inst{\ref{TBr}}
\and R. Bouwens\inst{\ref{RBo}}
\and O. Czoske\inst{\ref{OCz}}
\and M. Donahue\inst{\ref{MDo}}
\and H. Ford\inst{\ref{DLe}}
\and M. Girardi\inst{\ref{MGi},\ref{ABi}} 
\and L. Infante\inst{\ref{LIn}}
\and S. Jouvel\inst{\ref{SJo}}
\and D. Kelson\inst{\ref{DKe}}
\and A. Koekemoer\inst{\ref{MPo}}
\and U. Kuchner\inst{\ref{OCz}}
\and D. Lemze\inst{\ref{DLe}} 
\and M. Lombardi\inst{\ref{MLo}}
\and C. Maier\inst{\ref{OCz}}
\and E. Medezinski\inst{\ref{EMe}}
\and P. Melchior\inst{\ref{PMe}}
\and M. Meneghetti\inst{\ref{SEt},\ref{MMe}}
\and J. Merten\inst{\ref{LMo}}
\and A. Molino\inst{\ref{NBe}}
\and L. Moustakas\inst{\ref{LMo}}
\and V. Presotto\inst{\ref{MGi}}
\and R. Smit\inst{\ref{RBo}}
\and K. Umetsu\inst{\ref{KUm}}
}

\offprints{I. Balestra, \email{italobale@gmail.com}}

\institute{INAF - Osservatorio Astronomico di Trieste, Via G. B. Tiepolo 11, I-34131, Trieste, Italy\label{ABi} \and
INAF - Osservatorio Astronomico di Capodimonte, Via Moiariello 16 I-80131 Napoli, Italy\label{AMe} \and
INAF - Osservatorio Astronomico di Bologna, Via Ranzani 1, I-40127 Bologna, Italy\label{SEt} \and
Dipartimento di Fisica e Scienze della Terra, Universit\`a di Ferrara, Via Saragat 1, I-44122 Ferrara, Italy\label{PRo} \and
University Observatory Munich, Scheinerstrasse 1, D-81679 M\"unchen, Germany\label{SSe1} \and
Dark Cosmology Centre, Niels Bohr Institute, University of Copenhagen, Juliane Maries Vej 30, 2100 Copenhagen, Denmark\label{CGr} \and
Space Telescope Science Institute, 3700 San Martin Drive, Baltimore, MD 21218, USA\label{MPo} \and
INAF/IASF-Milano, via Bassini 15, 20133 Milano, Italy\label{MSc} \and
INAF - Osservatorio Astrofisico di Arcetri, Largo E. Fermi 5, 50125 Firenze, Italy\label{PTo} \and
Max-Planck-Institut f\"ur extraterrestrische Physik, Postfach 1312, Giessenbachstr., D-85741 Garching, Germany \label{SSe2} \and
Department of Physics and Astronomy, The Johns Hopkins University, 3400 North Charles Street, Baltimore, MD 21218, USA\label{DLe} \and
University of Vienna, Department of Astrophysics, T\"urkenschanzstr. 17, 1180 Wien, Austria\label{OCz} \and
Universit\"at Heidelberg, Philosophenweg 12, D-69120 Heidelberg, Germany\label{MBa} \and
Dipartimento di Fisica, Universit\`a degli Studi di Trieste, Via Tiepolo 11, I-34143 Trieste, Italy\label{MGi} \and
Instituto de Astrof\'{\i}sica de Andaluc\'{\i}a (CSIC), C/Camino Bajo de Hu\'etor 24, Granada 18008, Spain\label{NBe} \and
Department of Theoretical Physics, University of the Basque Country, P. O. Box 644, 48080 Bilbao, Spain\label{TBr} \and
Leiden Observatory, Leiden University, P. O. Box 9513,2300 RA Leiden, The Netherlands\label{RBo} \and
Department of Physics and Astronomy, Michigan State University, East Lansing, MI 48824, USA\label{MDo} \and
Observatories of the Carnegie Institution of Washington, Pasadena, CA 91 101, USA\label{DKe} \and
INFN - Bologna, Via Ranzani 1, I-40127 Bologna, Italy\label{MMe}  \and
Dipartimento di Fisica, Universit\`a degli Studi di Milano, via Celoria 16, I-20133 Milan, Italy\label{MLo} \and
Department of Physics and Astronomy, The Johns Hopkins University, 3400 North Charles Street, Baltimore, MD 21218, USA\label{EMe} \and
Institut de Ci\`encies de l'Espai (IEEC-CSIC), E-08193 Bellaterra (Barcelona), Spain\label{SJo}  \and
Departamento de Astronomia y Astrofisica, Pontificia Universidad Catolica de Chile, V. Mackenna 4860, Santiago 22, Chile\label{LIn} \and
Department of Physics, The Ohio State University, Columbus, OH, USA \label{PMe} \and
Jet Propulsion Laboratory, California Institute of Technology, 4800 Oak Grove Dr, Pasadena, CA 91109, USA\label{LMo} \and
Institute of Astronomy and Astrophysics, Academia Sinica, P. O. Box 23-141, Taipei 10617, Taiwan\label{KUm}
}

\date{Received 2013/ Accepted 2013}

\titlerunning{CLASH-VLT: Spectroscopic confirmation of z=6.11 lensed galaxy}
\authorrunning{I. Balestra et al.}

\abstract{...}
\abstract{We present VIsible Multi-Object Spectrograph (VIMOS) observations of a $z\sim6$ 
galaxy quintuply imaged by the Frontier Fields galaxy cluster RXC~J2248.7-4431 $(z=0.348)$. 
This sub-$L^*$, high-$z$ galaxy has been recently discovered by \citet{mon13} using dropout 
techniques with the 16-band HST photometry acquired as part of the Cluster Lensing And 
Supernova survey with Hubble (CLASH). Obtained as part of the CLASH-VLT survey, the 
VIMOS medium-resolution spectra of this source show a very faint continuum between 
$\sim8700\AA$ and $\sim9300\AA$ and a prominent emission line at 
8643$\AA$, which can be readily identified with Lyman-$\alpha$ at 
$z=6.110\pm0.002$. The emission line 
exhibits an asymmetric profile, with a more pronounced red wing. The rest-frame equivalent 
width of the line is $EW=79\pm10\,\AA$, relatively well constrained thanks to the 
detection of the UV continuum, which is rarely achieved for a sub-$L^*$ galaxy at this redshift. 
After correcting for magnification, the 
star formation rate (SFR) estimated from the Ly$\alpha$ line is 
SFR$($Ly$\alpha)=11\,$M$_{\odot}\,$yr$^{-1}$ and that estimated from the UV data is 
SFR$($UV$)=3\,$M$_{\odot}\,$yr$^{-1}$. We estimate that the effective radius of the source 
is $R_e\lesssim0.4$~kpc, which implies a star formation surface mass density 
$\Sigma_{SFR}>6\,$M$_{\odot}\,$yr$^{-1}\,$kpc$^{-2}$ and, using the Kennicutt-Schmidt relation, 
a gas surface mass density $\Sigma_{gas}>10^3\,$M$_{\odot}\,$pc$^{-2}$. Our results support 
the idea that this magnified, distant galaxy is a young and compact object 
with luminosity $0.4\,L^*$ at $z=6$, when the Universe was just 1~Gyr old, with 
a similar amount of mass in gas and stars. 
In the spirit of the Frontier Fields initiative, we also publish the redshifts of several 
multiply imaged sources and other background objects, which will help improving the 
strong-lensing model of this galaxy cluster. 
}

\keywords{Gravitational lensing: strong -- galaxies: high redshift.}

\maketitle


\section{Introduction}

Understanding the process of reionization of the intergalactic medium in the early 
Universe and the nature of the first galaxies responsible for that process are among the 
most important goals of modern cosmology \citep[e.g.][ and references therein]{rob10}.
In recent years, great progress has been made in our ability to detect galaxies at 
$z\gtrsim6$. To the greatest extent, this has been possible thanks to very deep 
observations with the Hubble Space Telescope (HST). The selection of optical dropouts 
in deep HST fields 
has led to the 
identification of samples that now reach hundreds of candidate Lyman-break galaxies (LBG) 
at $z\sim6-8$ \citep[e.g.][]{bou06,bou11,koe11,tre12,oes12,yan12,bra12,ell13,bra13}.

Acting as \textit{Cosmic Telescopes}, massive lensing galaxy clusters 
allow the identification 
of faint, otherwise undetected, high-$z$ galaxies. Despite the smaller volumes probed 
with gravitational lensing compared to those probed in deep fields, lensing has been 
successful in pushing the detection of distant galaxies one step farther, as testified 
by the recent discoveries of several $z\sim9-11$ candidates behind massive clusters 
\citep{bow12,zhe12,coe13}. All of these high-$z$ candidates have been identified 
by galaxy clusters observed as part of the Cluster Lensing And Supernova survey with 
Hubble \citep[CLASH;][]{pos12} Multi-Cycle Treasury program.

Ground-based spectroscopy of high-redshift sources in "blank" fields can be very 
challenging given that they are often 27th magnitude or fainter.  However, the flux 
magnifications due to gravitational lensing helps to bring these distant sources within 
reach of ground-based spectroscopy. Lensed observations reach significantly lower
luminosities \citep[e.g.][]{she12} than spectroscopic observations in deep 
fields \citep[e.g.][]{van11}.


The most prominent spectral feature in the UV rest-frame wavelengths probed by 
optical/NIR spectroscopy at $z\gtrsim6$ is the Lyman-$\alpha$ (Ly$\alpha$) emission line. 
The Ly$\alpha$ emission line is a resonant transition and suffers from radiative 
transfer effects, a property that further complicates the modeling of the escape 
fraction of Ly$\alpha$ photons from their host galaxies \citep{ver06,dij13}. 
The line itself is also an important diagnostic of the physical processes at work
\citep{gia96,van09}, since its strength and velocity profile depend on the instantaneous 
star formation rate, gas and dust content, metallicity, kinematics, and
geometry of the interstellar medium (e.g., clumpy, anisotropic).


The cluster RXC~J2248.7-4431 (z=0.348) was observed as part of the CLASH-VLT large 
spectroscopic program, which targets 14 CLASH clusters in the southern sky.
Five multiple images of a young galaxy, identified as $i$-dropouts, have been recently 
discovered from the CLASH HST imaging of this cluster \citep[see][ hereafter M13]{mon13}. 
Two of the brightest images (ID2 and ID3) also have Spitzer detections, which
additionally support the high-$z$ nature of the source (see M13).
The delensed UV luminosity of the magnified source, inferred from the strong-lensing 
analysis is $L_{1600}\sim0.4\,L^*_{1600}$ at $z=6$. The UV slope is steep ($\beta=-2.89\pm0.25$) 
and the inferred age ($<300$Myr), mass ($\sim10^8\,M_{\odot}$), and metallicity 
($Z<0.2\,Z_{\odot}$) are all consistent with that of a young galaxy (M13).

In this letter, we report on our VIMOS/VLT observations, which provide spectroscopic 
confirmation at $z=6.110$ for three of the high-redshift, quintuply imaged $i$-dropouts in 
RXC~J2248.7-4431, and we infer some of the physical properties of this young galaxy. 
In addition, we provide a list of redshift measurements for several other strong-lensing 
features and magnified background sources targeted in this cluster by our CLASH-VLT 
survey so far.
Errors are quoted to the $1\sigma$ confidence level, 
unless otherwise stated. 
We assume a cosmology with $\Omega_{\rm tot}, \Omega_M, \Omega_\Lambda = 1.0, 0.3,
0.7$ and $H_0 = 70$~km~s$^{-1}$~Mpc$^{-1}$.

\section{VIMOS observations and data reduction} \label{data}

\begin{figure}
\centering
\includegraphics[width=8.0 cm, angle=0]{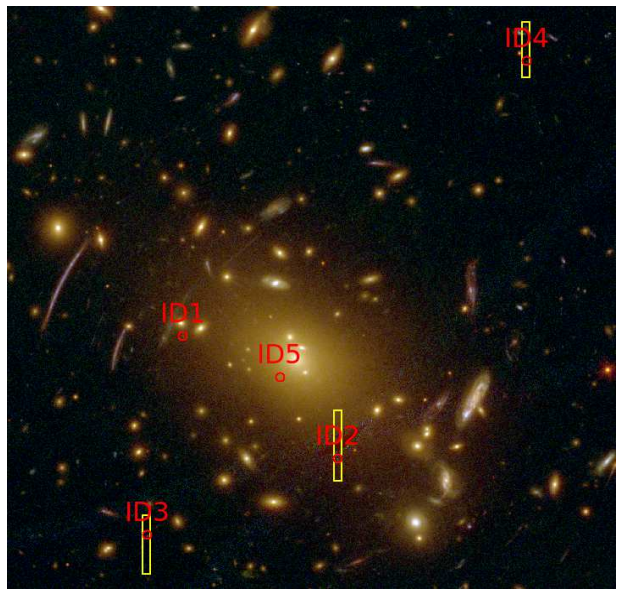}
\caption{HST color-composite image.  
The red circles (ID1-5) show the positions of the quintuply lensed dropout 
candidate at $z=6$, labeled as in M13.
The three VIMOS slits (yellow boxes) were placed on
ID2, ID3, and ID4 in four repeated 1hr-pointings. ID1 was not
targeted because it is too close to a bright cluster-member galaxy.}
\label{HST}
\end{figure}

The cluster RXC~J2248.7-4431 was observed in June-July 2013. 
The VIMOS data
were acquired using four separate pointings with one quadrant centered on the cluster 
core. Three of the five multiple images of the $i$-dropout candidate were targeted 
in four medium-resolution (MR) 1hr-pointings, providing us a total of 4~hr of integration 
time on each of the three images. 
The three multiple images targeted are displayed in Fig.~\ref{HST}. 
The identification 
numbers of the three targets are those used in M13. 
The masks were designed with $1''$-slits and the VIMOS spatial resolution is $0.205''/$pixel.
The MR grism and the GG475 filter were used. In this configuration the useable wavelength 
range is 4800-10\,000$\,\AA$, the resolution is $R=580$ (or $\sim13\AA$), and 
the dispersion is $2.55\AA/$pixel.

Data reduction was performed using the Vimos Interactive Pipeline 
Graphical Interface \citep[VIPGI; ][]{sco05}, which uses standard automated procedures 
to compute bias subtraction, flat-fielding, sky subtraction, and wavelength calibration.
The standard star EG-274 was used for flux calibration.

The seeing varied from pointing to pointing. The best seeing conditions 
($\sim0.6-0.7''$) were reached during pointing P1 and P3 (July 11 and 10), while 
they were poorer ($\sim1.0-1.1''$) during pointing P2 and P4 (July 6 and 9). 
As a consequence of the varying seeing conditions the signal-to-noise ratio (S/N) of 
the spectra varied significantly in the different pointings. 

\begin{figure}
\centering
\includegraphics[width=7.0 cm, angle=0]{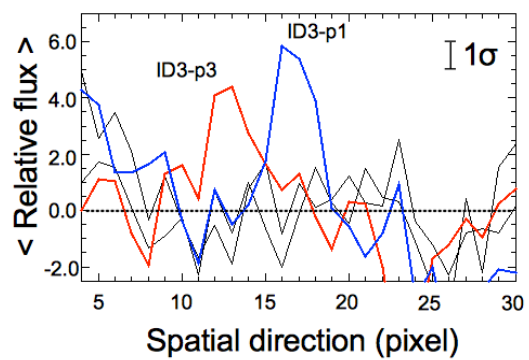}
\caption{Spatial profiles along the direction of the slit for the spectra of ID3, 
derived by collapsing counts in the dispersion direction between 
$9050\,\AA$ and $9280\,\AA$ for each pointing separately. P1 is plotted in 
\textit{blue}, P3 in \textit{red}, and P2 and P4 in \textit{black}. The 
continuum is clearly detected in P1 ($\sim5\sigma$) and P3 ($\sim3\sigma$). Spectra
are generally at slightly different pixel positions in different pointings.
}
\label{s2n}
\end{figure}

\begin{figure*}
\centering
\includegraphics[width=16.5 cm, angle=0]{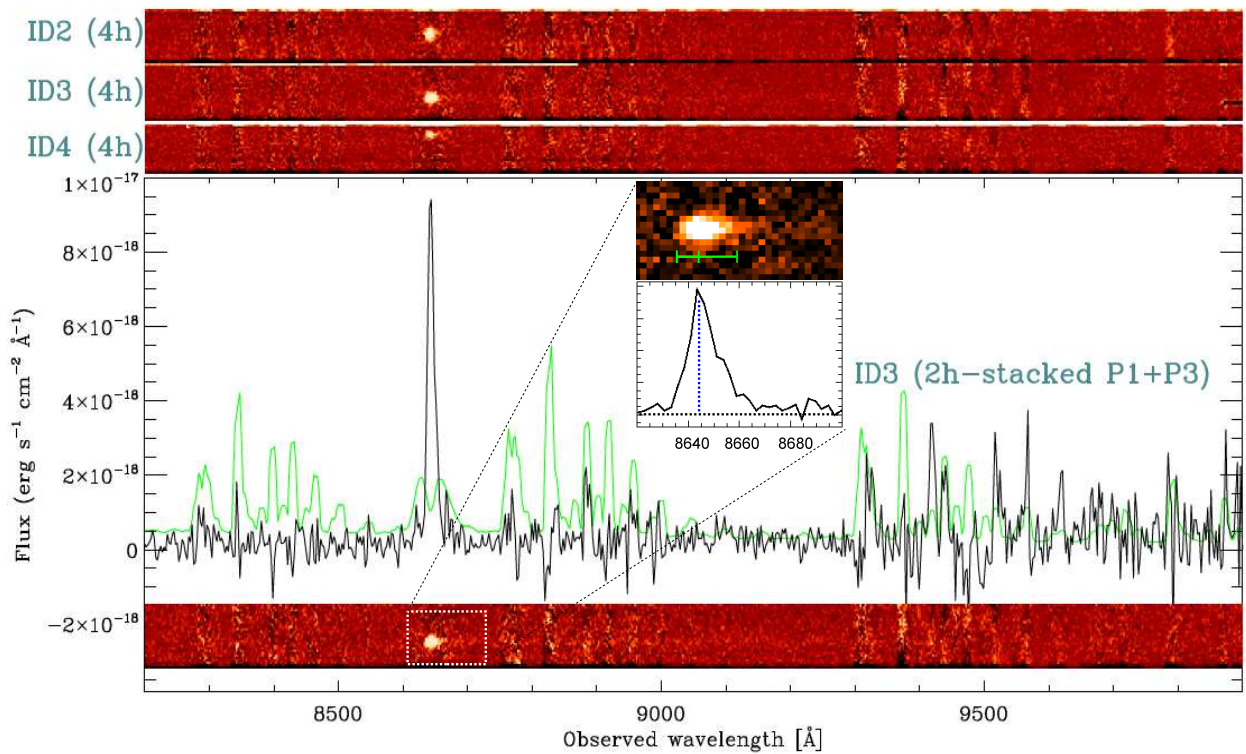}
\caption{VIMOS 1D spectrum of ID3 (2h-exposure, pointing P1+P3; \textit{black line}) and 
of the sky rescaled to arbitrary units (\textit{green line}). The lower inset shows 
the sum of the 2D spectra ($2\times$1hr exposure, P1+P3). The 4h-stacked 2D spectra of 
each multiple image (ID2, ID3, and ID4) are shown in the insets on the top. A zoomed 
version of the 1D and 2D spectra around the Ly$\alpha$ is shown in the central inset.
The vertical \textit{blue dotted} line marks the position of the peak of the line, 
which shows a clearly asymmetric profile. In the 2D spectrum, the \textit{green segments} 
mark the position of the peak of the flux and that of the 20\% flux level.}
\label{ID32h}
\end{figure*}


\section{Results}\label{results}

A strong emission line is detected at $8643\AA$ in each of the single 1hr-spectra (P1-P4) 
for all of the three multiple images targeted (ID2-ID4). If the emission line is 
identified to be Ly$\alpha$, the redshift of the source is $z=6.110$, which is consistent
both with the photo-z ($z_{phot}\simeq5.9$, see M13) and with the strong-lensing model 
predictions ($z_{lensing}\simeq6.0$, see M13). For the brightest image 
(ID3), in the two pointings with better seeing (P1 and P3) a faint continuum 
is also detected (at about 5$\sigma$) between $\sim8700\AA$ and $\sim9300\AA$ 
(see Fig.~\ref{s2n}). 
To increase the S/N, we tried to stack the four 1hr-spectra for each multiple 
image. However, the S/N is maximized when only the two spectra with the highest S/N 
(P1 and P3) are combined. The higher S/N in these two pointings is due to a
combination of better seeing and better transparency at the time of the observations. 
Fig.~\ref{ID32h} shows the sum of the 2D spectra for each
of the three multiple images (ID2, ID3, and ID4) and the highest S/N combined spectrum 
(2hr-stacked spectrum of ID3). Unfortunately, even in the stacked spectrum the S/N of 
the continuum is not sufficiently high to reliably identify any other UV absorption feature.
It is worth to recall that star-forming galaxies (and their stacks) with a large Ly$\alpha$
equivalent width ($EW>50\,\AA$) show very weak or even lack UV stellar and
interstellar absoption lines \citep[e.g.,][]{van09,bal10}, therefore it could be 
intrinsically difficult to detect them even in higher S/N spectra.
The lack of the N~V emission line suggests the absence of strong nuclear activity (AGN), 
corroborated by the lack of X-ray detection. However, since the 2$\sigma$ upper limit 
on the $0.5-10$~keV rest-frame luminosity is only $5\times10^{44}$~erg~s$^{-1}$ 
(obtained by stacking Chandra counts at the position of the three brightest magnified images), 
a low-luminosity AGN cannot be excluded with the current shallow X-ray.

To obtain a more precise measurement of the continuum flux and, hence, a more 
reliable estimate of the $EW$ of the Ly$\alpha$, we used 
only the highest S/N spectrum (ID3 2hr-stacked spectrum). The integrated flux of the 
line is $1.63\times10^{-16}$~erg~s$^{-1}$~cm$^{-2}$ and the continuum has a flux of 
$2.9\times10^{-19}$~erg~s$^{-1}$~cm$^{-2}\,\AA^{-1}$ in the wavelength range 
$\sim9000-9300\AA$. This yields an equivalent width of $EW=79\pm10\AA$ for the 
Ly$\alpha$, where the error is estimated from the statistical error on the flux of the 
continuum.

The line profile is asymmetric with a more pronounced red wing (see inset of 
Fig.~\ref{ID32h}). The asymmetry of the line is typically observed at these redshifts and 
in this case is mainly due to the intergalactic medium, which at $z\sim6$ can completely 
absorb half of the line \citep[e.g. ][]{lau09}. Other processes can shape the line 
profile, but the spectral resolution of our observational setup prevents us from 
separating possible secondary features related to internal kinematics (e.g., outflows).


We derived the star formation rate (SFR), following Kennicutt (1998), by converting
the observed nebular (Ly$\alpha$) and stellar ultraviolet (UV $1600\AA$) fluxes,
without correction for dust attenuation. After correcting for the magnification factor
($\mu=6$), the SFR is SFR(Ly$\alpha)=11\,$M$_{\odot}\,$yr$^{-1}$ and 
SFR(UV$)=3\,$M$_{\odot}\,$yr$^{-1}$, respectively. The intrinsic SFR(Ly$\alpha$) can be 
even higher (possibly double) since the IGM most probably attenuates $\sim50\%$ of the 
line flux. Interestingly, the SFR(Ly$\alpha$) is higher, or at the most similar to that 
derived from the UV continuum. As discussed in \citet{ver08}, 
this is indicative of very low or negligible dust attenuation coupled with
very young stellar populations and recent onset of star formation activity 
(consistent with recent results by M13). Given the stellar mass derived from 
SED fitting ($\sim2\times10^8\,$M$_{\odot}$, see M13), the specific 
star formation rate is sSFR(Ly$\alpha)=55\,$Gyr$^{-1}$ (or sSFR(UV$)=15\,$Gyr$^{-1}$).

To estimate the size of the source we measured the effective radii of ID2, ID3, and ID4 
from the HST images and the magnification factors obtained from the lensing model by M13.
We estimate that the effective radius of the source is $R_e\lesssim0.4$~kpc. 
Given our estimate of the size and the value of SFR(UV), we can compute the SFR 
surface mass density $\Sigma_{SFR}>6\,$M$_{\odot}\,$yr$^{-1}\,$kpc$^{-2}$ 
and we can use the Kennicutt-Schmidt relation to derive the gas surface mass density 
$\Sigma_{gas}>10^3\,$M$_{\odot}\,$pc$^{-2}$.
This indicates that the stellar-over-gas mass ratio is close to
unity, in line with theoretical expectations for small mass 
($\lesssim 10^{9}\,$M$_{\odot}$) and low SFR objects at these redshifts 
\citep[see][]{calu08}.

\subsection{Redshifts of arcs and other background objects}

As part of our CLASH-VLT survey, we targeted several multiply imaged 
sources and candidate high-$z$ objects. In Table~\ref{arcs}, we report the 
spectroscopic redshifts obtained so far for this cluster also with the low-resolution blue 
grism. We confirm 10 of the 14 
multiple image systems used by M13. These measurements confirm the high accuracy of 
the photometric redshifts and additionally support the strong-lensing model 
($z_{lens}\simeq z_{spec}$).

\begin{table}
\caption{Redshifts of 10 multiple image systems and other background objects measured 
in the core of RXC~J2248.7-4431 to date.}
\begin{center}
\begin{tabular}{c c c l l}
\hline
\hline
 RA          &  Dec          & $z_{spec}$ & $z_{phot}$ &  $z_{lens}$ \\
(1)          & (2)           & (3)        & (4)        & (5)         \\
\hline
 22:48:43.45 &   -44:32:04.6 &  6.110 & $5.87^{+0.03}_{-0.02}$ & 6.0 (ID2)   \\
 22:48:45.81 &   -44:32:14.8 &  6.110 & $6.01^{+0.03}_{-0.06}$ & 6.0 (ID3)   \\
 22:48:41.11 &   -44:31:11.4 &  6.110 & $5.95^{+0.06}_{-0.08}$ & 6.0 (ID4)   \\
 22:48:47.00 &   -44:31:44.0 &  1.229 & $1.26^{+0.03}_{-0.06}$ & 1.19 (1.1a)  \\
 22:48:44.75 &   -44:31:16.3 &  1.229 & $1.22^{+0.05}_{-0.02}$ & 1.19 (1.1c)  \\
 22:48:46.22 &   -44:31:50.6 &  1.260 & $1.23^{+0.06}_{-0.06}$ & 1.26 (3a)    \\
 22:48:45.08 &   -44:31:38.4 &  1.398 & $1.53^{+0.02}_{-0.02}$ & 1.40 (4b)    \\
 22:48:43.01 &   -44:31:24.9 &  1.398 & $1.11^{+0.02}_{-0.02}$ & 1.40 (4c)    \\
 22:48:45.22 &   -44:32:24.0 &  1.429 & $1.17^{+0.10}_{-0.09}$ & 1.46 (6a)    \\
 22:48:41.56 &   -44:32:23.9 &  3.110 & $3.03^{+0.05}_{-0.05}$ & 2.97 (11b) \\
 22:48:41.38 &   -44:32:28.4 &  1.270 & $1.18^{+0.04}_{-0.07}$ & $-$  \\
 22:48:46.92 &   -44:32:49.4 &  1.353 & $1.18^{+0.02}_{-0.03}$ & $-$  \\
 22:48:39.02 &   -44:32:34.6 &  3.242 & $3.44^{+0.07}_{-0.07}$ & $-$  \\
 22:48:49.28 &   -44:30:55.8 &  3.542 & $3.55^{+0.19}_{-0.07}$ & $-$  \\
 22:48:44.23 &   -44:31:31.0 &  0.730 & $0.70^{+0.03}_{-0.03}$ & $-$  \\
\hline
\end{tabular}
\end{center}
\begin{tiny}\textbf{Notes.} (1-2) J2000 coordinates, (3) spectroscopic redshift, (4-5) 
photometric and lensing-model predicted redshift (and IDs) from M13.                                       
\end{tiny}  
\label{arcs}
\end{table}


\section{Discussion}\label{dis}

We presented VIMOS/VLT observations providing a spectroscopic confirmation at $z=6.110$ 
for a galaxy quintuply imaged by the Frontier Fields cluster RXC~J2248.7-4431. 

The VIMOS spectra clearly show a strong Ly$\alpha$ and a relatively faint, but 
significantly detected ($5\sigma$), continuum.
This is a rare case where the detection of the continuum can be 
achieved in galaxies with delensed magnitudes of $\sim27$. Remarkably, this 
was achieved in only 1 hr of exposure.
Our results, together with those recently presented by M13, suggest 
that this magnified, distant galaxy is a young ($<300\,$Myr, $\sim10^8\,$M$_{\odot}$, 
and $Z<0.2\,Z_{\odot}$) and compact ($\lesssim0.4$~kpc)
object fainter than $L^*$ at $z=6$, with a similar amount of mass in gas and stars.
We can rule out the presence of a strong AGN because of the absence of N~V emission 
line in the VIMOS spectra and because of the Chandra upper limit on the X-ray luminosity 
($<5\times10^{44}$~erg~s$^{-1}$).

While this paper was finalized, \citet{boo13} very recently reported the detection of 
a submillimeter source with APEX/LABOCA at $870\mu$m, which they tentatively associated
with the $z\sim6$ multiple system, or alternatively with a second interacting galaxy, 
or even with the SZ signal from the ICM. The association of the submm source with the 
$z\sim6$ multiply imaged galaxy remains extremely uncertain due to the poor resolution 
of submm observations and to foreground contamination.
The SZ explanation or the two-source hypothesis seem more plausible, because in the
single-source scenario it would be very difficult to explain how
the dust emission in the submillimeter ($A_v\sim1.5$) can be reconciled
with the extremely steep UV spectral slope derived from
CLASH photometry ($\beta \sim -2.9$, see M13) and the prominent
Ly$\alpha$ emission, $EW\simeq80\AA$. Both features typically indicate
very low dust attenuation \citep{hay11,deb12,ate13,lau13}. 
Therefore, we interpret the HST $z=6.110$ source as a young and compact object 
with low dust content in an early phase of evolution, when the Universe was just 
1~Gyr old. High spatial-resolution submillimiter observations with ALMA will be 
extremely useful to reliably constrain the SFR and the molecular gas content
in this magnified high-$z$ system.

\begin{acknowledgements}
We thank the anonymous referee for the valuable comments and suggestions.
We acknowledge partial support by the DFG Cluster of Excellence Origin Structure of the 
Universe. AZ is supported by contract research ``Internationale Spitzenforschung II/2-6'' 
of the Baden-W\"urttemberg Stiftung.
\end{acknowledgements}

\bibliographystyle{aa}
\bibliography{sbs}



\end{document}